\documentclass[journal]{IEEEtran}
\pdfoutput=1
\usepackage[noadjust]{cite}
\usepackage{graphicx}
\usepackage{amsmath}
\usepackage{amssymb}
\usepackage{multirow}
\usepackage{subfigure}
\usepackage[lined,boxed, ruled]{algorithm2e}
\usepackage{color}

\begin{document}

\title{Dense Dilated Network with Probability Regularized Walk for Vessel Detection}
\author{Lei~Mou,~Li~Chen,\thanks{This work was supported in part by National Natural Science Foundation of China (61773297  and 61601029), Open Fund of Hubei Province Key Laboratory of Intelligent Information Processing and Real-Time Industrial System (2016ZNSS01A), Grant of Ningbo 3315 Innovation Team and also Zhejiang Provincial Natural Science Foundation (LZ19F010001) \emph{(Corresponding author: Jun Cheng and Li Chen.)}}
	\thanks{L. Mou is with School of Computer Science and Technology, Wuhan University of Science and Technology, Wuhan 430081, China,  Hubei Province Key Laboratory of Intelligent Information Processing and Real-time Industrial System, Wuhan University of Science and Technology, Wuhan 430081, China and also Cixi Institute of Biomedical Engineering, Chinese Academy of Sciences, Zhejiang 315201, China. E-mail: moulei@nimte.ac.cn.}
	\thanks{L. Chen is with School of Computer Science and Technology, Wuhan University of Science and Technology, Wuhan 430081, China and  Hubei Province Key Laboratory of Intelligent Information Processing and Real-time Industrial System, Wuhan University of Science and Technology, Wuhan 430081, China. E-mail: chenli@wust.edu.cn.}\and Jun Cheng,  Zaiwang~Gu,  Yitian Zhao 	and Jiang Liu
		\thanks{J.~Cheng is with Cixi Institute of Biomedical Engineering, Chinese Academy of Sciences, Zhejiang 315201, China and also with UBTech Research, UBTech Robotics Corp Ltd, Guangdong 518055, China. Email: sam.j.cheng@gmail.com.}
			\thanks{Z. Gu is with the Department
		of Computer Science and Engineering, Southern University of Science and Technology, Guangdong 518055, China Email: guzaiwang@gmail.com.}
	 \thanks{Y. Zhao is with the Cixi Institute of Biomedical Engineering, Ningbo Institute of Industrial Technology, Chinese Academy of Sciences, Zhejiang 315201, China. E-mail: yitian.zhao@nimte.ac.cn.}
  \thanks{J. Liu is with the Department of Computer Science and Engineering,
 	Southern University of Science and Technology, Guangdong 518055, China
 	and also with the Cixi Institute of Biomedical Engineering, Chinese Academy
 	of Sciences, Zhejiang 315201, China. Email: liuj@sustech.edu.cn.}
}

\maketitle 
\begin{abstract}
  The detection of retinal vessel is of great importance in the diagnosis and treatment of many ocular diseases. Many methods have been proposed for vessel detection. However, most of the algorithms neglect the connectivity of the vessels, which plays an important role in the diagnosis. In this paper, we propose a novel method for retinal vessel detection. The proposed method includes a dense dilated network to get an initial detection of the vessels and a probability regularized walk algorithm to address the fracture issue in the initial detection. The dense dilated network integrates newly proposed dense dilated feature extraction blocks into an encoder-decoder structure to extract and accumulate features at different scales. A multi-scale Dice loss function is adopted to train the network. To improve the connectivity of the segmented vessels, we also introduce a probability regularized walk algorithm to connect the broken vessels. The proposed method has been applied on three public data sets: DRIVE, STARE and CHASE\_DB1. The results show that the proposed method outperforms the state-of-the-art methods in accuracy, sensitivity, specificity and also area under receiver operating characteristic curve.
\end{abstract}
\begin{IEEEkeywords}
Vessel segmentation, encoder-decoder, deep learning, regularized walk, vessel reconnection
\end{IEEEkeywords}
%%%%%%%%%%%%%%%%%%%%%%%%%%%%%%%%%%%%%%%%%%%%%%%%%%
\begin{figure*}
	\centering
	\includegraphics[width=0.98\textwidth]{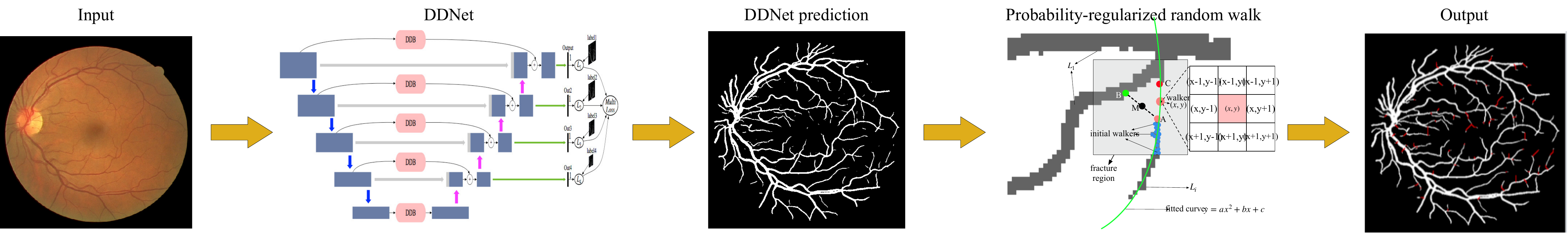}
	\caption{The overall flow chart of the proposed method. It includes a dense dilated convolutional network (DDNet) and a regularized walk algorithm. The DDNet is based on encoder-decoder structure, which is integrated with newly proposed dense dilated blocks (DDB)  between the encoder and decoder to compute an initial map of the vessels. Then a probability regularized walk is used to improve the results from the initial map.}
	\label{fig:architecture}  
\end{figure*}
%%%%%%%%%%%%%%%%%%%%%%%%%%%%%%%%%%%%%%%%%%%%%%%%%%%%
\section{{Introduction}}
Retinal vessels are commonly analyzed in the diagnosis and treatment of various ocular diseases such as age-related macular degeneration, diabetic retinopathy, glaucoma, hypertension, arteriosclerosis and multiple sclerosis \cite{venkataraman2010vascular} \cite{flammer2002impact} \cite{vermeer2004model}. Clinicians have found that the structure of retinal blood vessels is associated with diabetes \cite{jelinek2009automated} \cite{sinthanayothin1999automated} which affects the lining of the blood vessels in eyes. Poor circulation in the retinal vessels often compounds these problems by causing the generation of fragile new vessels. Further, vessel segmentation is also the basis for subsequent retinal artery/venous vessel classification task, which is of direct assistance to the ophthalmologist in terms of diagnosis and treatment of eye disease \cite{zhao2018retinal,zhao2019retinal}. Therefore, it is important for the clinicians to segment and analyze the retinal vessels, especially the tiny ones. However, manual demarcation of retinal vessels is a costly and time-consuming task. In recent years, extensive work has been done to segment the vessel automatically. Li \textit{et al.} used the  {Hessian} matrix and random walks to segment retinal vessel \cite{li2015automated}. Nguyen \textit{et al.} utilized a multi-scale line detection scheme for vessel segmentation \cite{nguyen2013effective}. Bankhead \textit{et al.} developed a fast retinal vessel detection method by using wavelets \cite{daubechies1992ten} and edge location refinement \cite{bankhead2012fast}. Soares \textit{et al.} used supervised classification to apply wavelets to 2-D retinal blood vessel segmentation \cite{soares2006retinal}. L\"{a}th\'{e}n \textit{et al.} proposed a quadrature filtering method with   multi-scale analysis for retinal vessel segmentation \cite{lathen2010blood}. Zhao \textit{et al.} \cite{zhao2017saliency} further proposed a vessel segmentation model based on saliency guided infinite perimeter active contour. A more detailed review of traditional vessel detection methods can be found in \cite{zhao2015automated} \cite{zhao2018automatic}. A common limitation of these traditional vessel detection algorithms is that they often require some manually determined parameters tuned based on local data set, which makes the algorithms less scalable.

With the rapid development of deep learning, many algorithms based on deep learning have been proposed for retinal blood vessel detection. Fu \textit{et al.} proposed a loss function based on multi-scale output combined with conditional random field (CRF) to improve the detection of vessels in retinal images with lesions \cite{fu2016deepvessel}.  Liskowski \textit{et al.} utilized global contrast normalization and zero-phase component analyzing techniques to enhance training samples for better blood vessel detection \cite{liskowski2016segmenting}. 
Recently, Gu \textit{et al.} proposed context encoder which utilizes dense atrous convolution and residual multi-kernel pooling for biomedical image segmentation \cite{Gu2019}. Zhang \textit{et al.} \cite{ET-Net} proposed ET-Net which  {focuses} on the guidance of edge information to segment optic discs, retinal vessels and lung organs. An efficient AG-Net \cite{etnet} has also been proposed to take attention mechanism into CNN to detect vessels in fundus.
Compared with the traditional methods, deep learning-based algorithms have better scalability and achieve better accuracy. The emergence of the encoder-decoder structure enables the network to extract deeper features while extracting the context information of the vessels \cite{unet} \cite{segnet}, thereby obtaining more accurate segmentation. In order to achieve higher segmentation accuracy, recent semantic segmentation algorithms \cite{lin2017refinenet}, \cite{bilinski2018dense}, \cite{yang2018denseaspp} adopted the state-of-the-art classification structures such as ResNet \cite{resnet}, ResNeXt \cite{resnext} and  {DenseNet} \cite{densenet}. In addition, multi-scale fusion strategy has been used to train the network for feature extraction at different resolutions \cite{zhao2017icnet} \cite{fu2018joint} \cite{chen2018deeplab}. In convolutional networks, pooling operations are used to obtain a larger receptive field, which leads to some loss of  {low-level} information. To avoid this issue, Yu and Koltun \cite{yu2015multi} designed a scheme called dilated (or atrous) convolution which is able to increase the receptive field without pooling. Prior  {works show} that atrous convolution can be combined with batch normalization to achieve more accurate results \cite{chen2017rethinking} \cite{wang2018understanding}. Recently, generative adversarial network  \cite{zhang2019skrgan} has also been proposed to improve vessel detetion.  

Although these deep learning-based methods show better performance in vessel detection \cite{fu2016deepvessel} \cite{soares2006retinal} \cite{liskowski2016segmenting}, they  {always} fail to detect small or tiny vessels, which lead to the fracture of the vessels. As a result, these methods often have   poor performance on the detection of tiny blood vessels. Clinically, tiny vessels provide an indispensable reference for the diagnosis of diseases like neovascular diseases. The  retinal vessels  {provide} a unique opportunity for cerebral small vessel disease study.  { For} example, subjects with smaller retinal arteriolar-to-venular ratio tended to have more white matter lesions \cite{ikram2005retinal}. Vessel connectivity has a major impact on the screening of vascular lesions. Therefore, improving the detection of tiny vessels in fundus images and effectively connecting fractured segmentations are crucial for ocular disease detection.

In this paper, we propose a dense dilated network (DDNet) and a probability regularized walk (PRW) algorithm for vessel detection. Fig. \ref{fig:architecture} shows the overall architecture of the proposed method, which includes the DDNet to  {obtain} an initial map of retinal vessels and the PRW algorithm to connect the broken vessels in the initial map.
In the vessel reconnection, we propose to integrate the probability output of the deep learning and the local vessel  {directions} into the regularized walk algorithm for fractured vessel reconnection.  
 Our main contributions are as follows:
\begin{enumerate}
\item  We propose a dense dilate network for retinal vessel detection. 
\item We propose a probability regularized walk algorithm to connect broken vessels from the initial vessel detection. It integrates the output of the dense dilated network and the local vessel directions into the regularized walk.
\item Experimental results show that the proposed method outperforms other methods for retinal vessel detection.
\end{enumerate}

The rest of the paper is organized as follows. Section \ref{relatedworks} reviews the related work. Section \ref{method} introduces our  {method} including DDNet and  {PRW}. In Section \ref{experiments}, we introduce the experimental results. The last section concludes the paper.
%=================================================================================================
%%%%%%%%%%%%%%%%%%%%%%%%%%%%%%%%%%%%%%%%%%%%%%%%
\begin{figure*}
\centering
	\includegraphics[width=\textwidth]{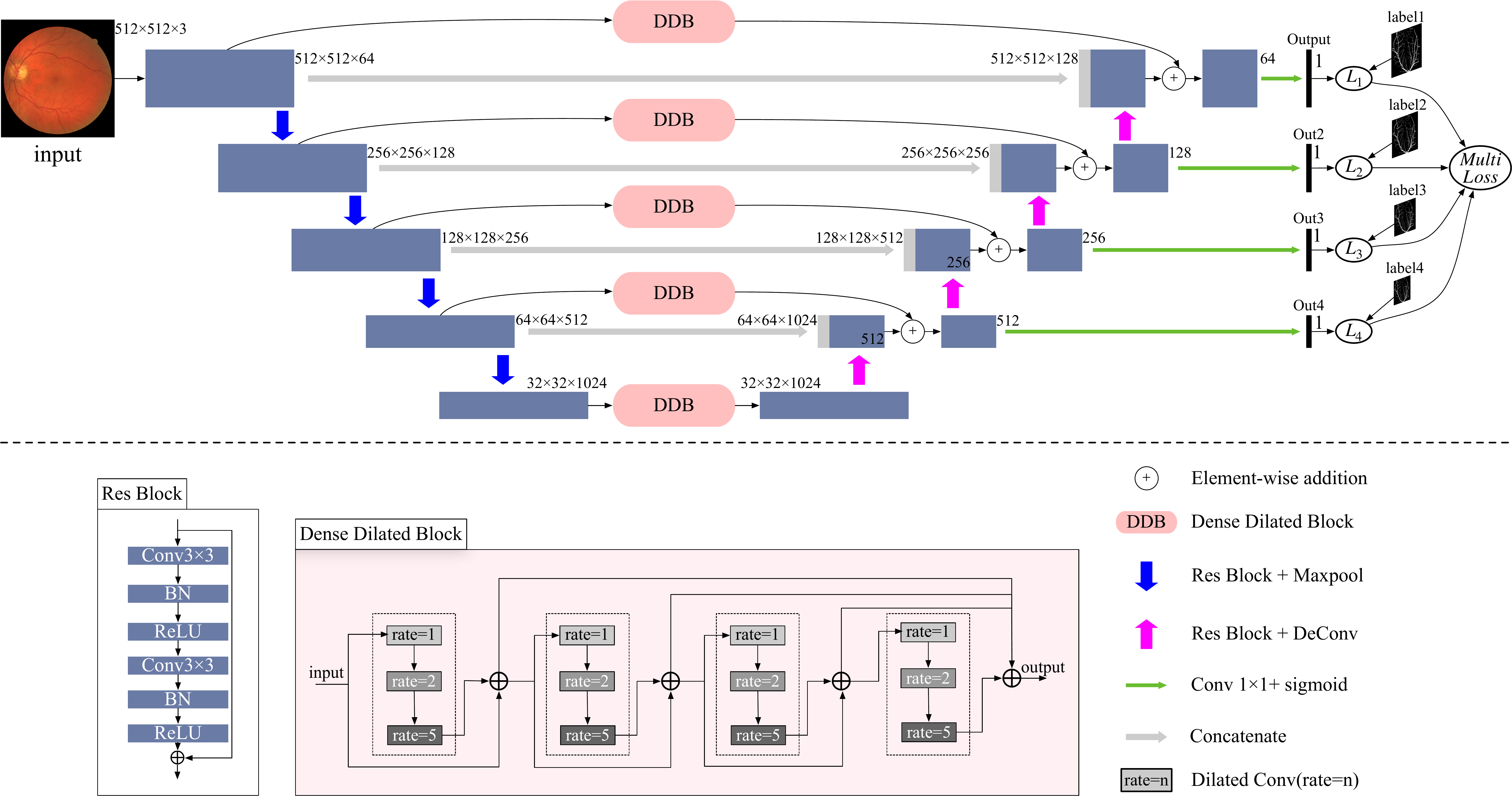}
	\caption{Diagram of the DDNet structure. In DDNet, there are three modules in total. The first module is the encoder-decoder infrastructure in which residual blocks are used to extract image features for vessel segmentation. The second module is  {dense} dilated block. In this module, we compute features in different scales and combine them  for segmenting blood vessels. The last module is the multi-scale dice loss.}
\label{ddnet}
\end{figure*}
%%%%%%%%%%%%%%%%%%%%%%%%%%%%%%%%%%%%%%%%%%%%%%%%%
\section{{Related Works}}
\label{relatedworks}
\subsection{{Encoder-Decoder}}
The encoder-decoder structure is one of the most popular frameworks for semantic segmentation. A notable characteristic of the encoder-decoder framework is that it is an end-to-end learning algorithm. Many segmentation methods are based on the encoder-decoder structure, such as SegNet \cite{segnet} and U-Net \cite{unet}. The encoder extracts the shallow and deep features of the images, which can be implemented in a variety of ways. Many different  {modules} such as VGG16 \cite{vgg16}, GoogLeNet \cite{googlenet} and ResNet \cite{resnet} can be used in the encoder.  In order to preserve the spatial information, Yu \textit{et al.} \cite{yu2017dilated} proposed a dilated residual network to extract features in the encoder. Zhao \textit{et al.} \cite{zhao2017icnet} proposed the image cascade network (ICNet) by utilizing the advantages of image pyramid in PSPNet \cite{zhao2017pyramid} to achieve reasonable performance in computation and segmentation of high-resolution images. In a recent work, Zhou \textit{et al.} \cite{zhou2018d} proposed a hybrid block of dilated convolutions with different rates between the encoder and the decoder to  {extract better} features. Pohlen \textit{et al.} proposed the full-resolution residual network \cite{pohlen2017fullresolution} model to enhance the final classification performance by combining different stages of information. The internal structure of the decoder is also important to improve the segmentation performance of the model. Recent work shows that   unpooling and deconvolution operations   in the decoder are beneficial   {for certain} segmentation tasks   \cite{noh2015learning}.

\subsection{{Vessel Detection}}
Retinal vessels are widely used in the diagnosis of various ocular diseases. In recent years, a variety of methods have been proposed to detect retinal vessels from fundus images, including unsupervised approaches \cite{marin2011new} and supervised approaches \cite{orlando2014learning}. The earlier traditional method \cite{ricci2007retinal} used support vector machines. Zhao \textit{et al.} \cite{zhao2015automated} \cite{zhao2018automatic} proposed infinite perimeter active contour model with hybrid region information and a weighted symmetry filter to detect vessels. Some other methods used a variety of filters to enhance the blood vessels \cite{jerman2016enhancement} \cite{lathen2010blood} \cite{azzopardi2015trainable}.  Marin \textit{et al.} \cite{marin2011new} employed a gray-level vector and moment invariant features using a neural network. Staal \textit{et al.} \cite{drive} detected vessels by detecting and grouping the ridge features. Soares \textit{et al.} \cite{soares2006retinal} used a multi-scale Gabor transform to extract features. Zhou \textit{et al.} \cite{zhou2017improving} proposed to learn more discriminative  {features} by improving dense CRF.
Zhang \textit{et al.} \cite{7530915} proposed orientation scores for robust vessel detection.
 In the field of deep learning, encoder-decoder structures \cite{unet} have shown to be promising for vessel detection. Maninis \textit{et al.} \cite{Maninis16} proposed a multi-task structure for both vessel detection and optic disc segmentation. Fu \textit{et al.} \cite{fu2016deepvessel} proposed DeepVessel model to improve the sensitivity of segmentation by introducing side-output layers and CRF.  Recently, Zhang \textit{et al.} \cite{zhang2018deep} introduced an edge-aware mechanism to convert the task into a multi-class task. Oliveira \textit{et al.} \cite{oliveira2018retinal} and Wu \textit{et al.} \cite{wu2018multiscale} proposed to segment retinal vessels based on conventional FCN and multi-scale architecture, respectively. Compared with \cite{zhang2018deep}–\cite{wu2018multiscale}, we have taken the vessel connectivity into consideration using probability regularized walk,  {however \cite{zhang2018deep}-\cite{wu2018multiscale} and other methods have not taken this into account.} In \cite{zhang2018deep}, the method has considered the importance of thin vessels by distinguishing the thin  {vessels} from thick vessels to force the networks to treat background pixels. However, this is different from our method to improve connectivity.

\subsection{{Vessel Reconnection}}
Although many methods have been proposed for  {vessel} detection, few of them deal with the vessel connection in the detection. Favali \textit{et al.} \cite{favali2016analysis} matched vessel segments by utilizing the vessel connectivity based on spectral clustering. Joshi \textit{et al.} \cite{joshi2011identification} presented a method for automatic identification and reconnection of broken vessels in the segmentation. Zhang \textit{et al.} \cite{zhang2018reconnection} connected the broken curvilinear structures using cortically inspired completion in ocular images.
% None of these methods employed the probabilities predicted by the deep learning model when connecting the broken vessels.

Random walk is a technology that has applications in many fields. Some methods employed random  {walk} for object segmentation \cite{meila2001learning} \cite{grady2006random} \cite{cheng2011connectedness}. In recent years, random  {walk has} been widely used in vessel segmentation. Li \textit{et al.} achieved considerable results using the hessian-based filter and random walk for vessel segmentation in \cite{li2015automated}. M'hiri \textit{et al.} \cite{m2013vesselwalker} applied the similar method to coronary arteries. Zhu \textit{et al.} \cite{zhu2013random} segmented the vessel by using random  {walk} with adaptive cylinder flux.

\section{Dense Dilated Network} 
\label{method}
Our proposed DDNet is an end-to-end deep neural network that mainly contains three parts. As shown in Fig. \ref{ddnet}, the first part is the backbone encoder-decoder structure for feature extraction and up-sampling. The second part is the dense dilated block (DDB), which has been integrated with the backbone. The design details of DDB are given in \ref{densedilatedblock}. The last part is the computation of the loss.

\subsection{Backbone Encoder-Decoder} 
\label{networkarchitecture}
Since the encoder-decoder structure has shown promising results for semantic segmentation, we use the encoder-decoder structure as the backbone of our DDNet. Following the work of Zhang \textit{et al.} \cite{zhang2018road}, we replace the original blocks in the U-Net with ResNet \cite{resnet} blocks to extract feature in our method. Our results  {show} that the replacement with ResNet blocks improves the semantic  {segmentation} as it enables feature reuse and mitigates the  {gradient} vanishing problem. 
% It should be noted that other blocks can be used flexibly as well, such as ResNeXt \cite{resnext} and Xception \cite{chollet2017xception}.

\subsection{Dense Dilated Block}
\label{densedilatedblock}
Recent studies in deep learning show that increasing the receptive field of the convolutional  {filter is} important to  {salient feature extraction} with more spatial information \cite{szegedy2017inception}. Earlier methods often use pooling layers and  {increase} kernel size to  {enlarge} the receptive field. However, pooling  {operation will potentially} lead to the loss of some spatial information while the increased kernel size requires more computation. To overcome this limitation, dilated convolutions are adopted to replace the original convolutions \cite{xie2015holistically}. Dilated convolution is a useful approach to adjust receptive  {field} of feature points without pooling or increasing the kernel size. Meanwhile, convolutional layers in a dense mode  {are} found to support feature reuse and make the network more efficient \cite{densenet}.

%%%%%%%%%%%%%%%%%%%%%%%%%%%%%%%%%%%%%%%%%%%
\begin{figure}[t]
\centering
	\includegraphics[width=0.48\textwidth]{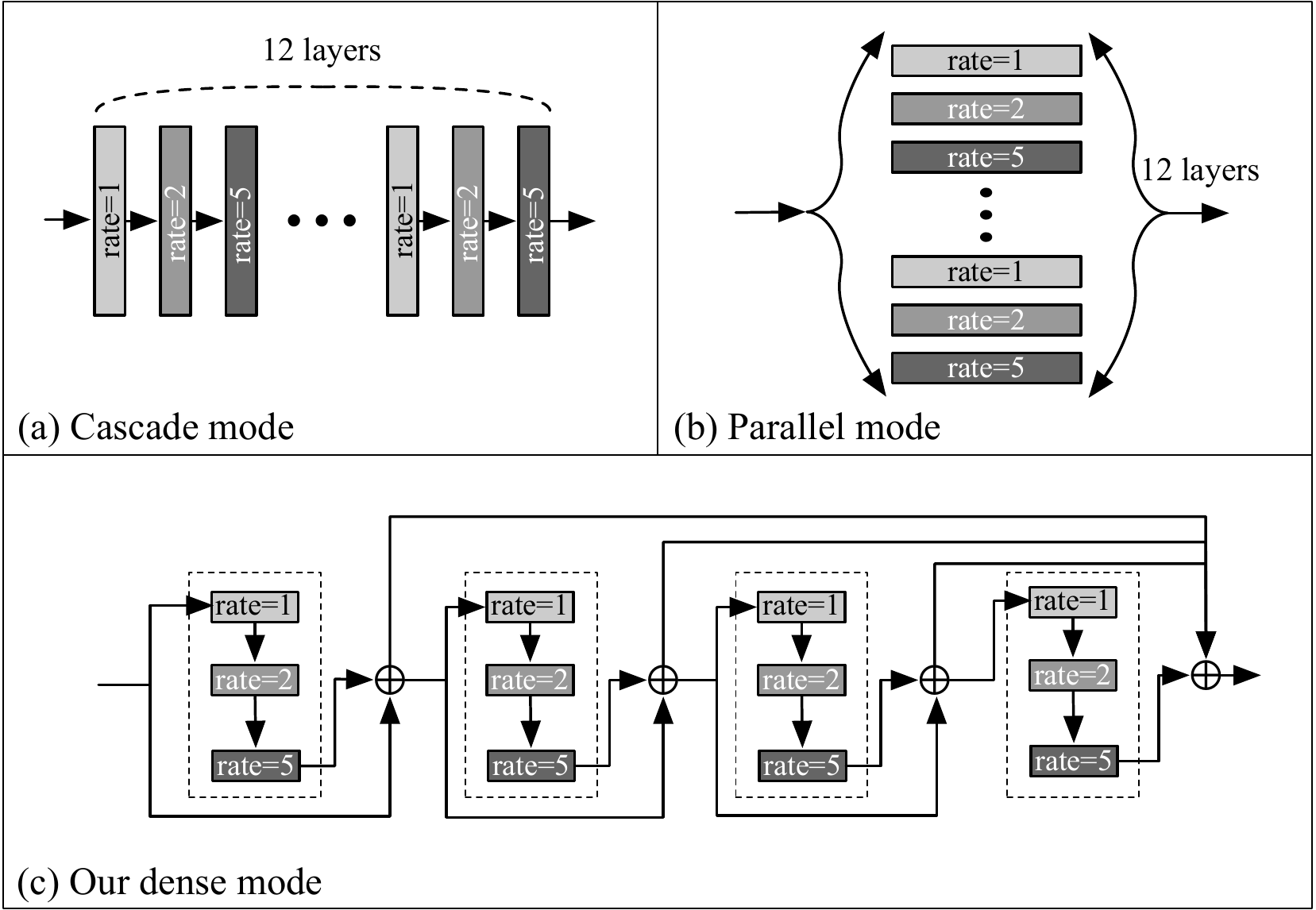}
	\caption{Diagram of different modes of dilated convolution. (a) cascade mode, and (b) parallel mode. (c) dense mode.}
	\label{dilatedconv}
\end{figure}
%%%%%%%%%%%%%%%%%%%%%%%%%%%%%%%%%%%%%%%%%%%%%

Motivated from the above  {observation}, we propose dense dilated block (DDB) to organize atrous convolutions in a dense form. As shown in Fig. \ref{dilatedconv}(c), the proposed  {DDB} contains four major blocks and each major block has three dilated convolutions. Wang \textit{et al.} \cite{wang2018understanding} have successfully demonstrated the benefits of the dilated convolution for the segmentation. In our paper, we follow the work to use three dilated convolutions with rate 1, 2, and 5. Our method is different    from existing   cascade mode \cite{yu2015multi} and parallel mode \cite{chen2018encoder}. Fig. \ref{dilatedconv} compares the three different modes. The benefits of  {this} organization are twofold: first, it reuses the features of the early layers and makes the network more efficient; second, we adopt dilated convolution in each layer of DDB to extract deeper features  {that avoiding} the reduction of the feature map resolution when increasing the receptive field. The depth-concatenate in our dense mode establishes the connection between different layers and  {makes} full use of the features  {in different scales}.  
\subsection{Multi-Scale Dice Loss}
\label{mlloss}
The loss function is an important factor in deep learning. Recent work \cite{xie2015holistically} demonstrates that the multi-scale loss effectively assists the training of the models. In the classification task, Li \textit{et al.} \cite{li2017person} combined the loss of local and global features to improve the    model. In the semantic segmentation task, Fu \textit{et al.}\cite{fu2018joint} up-sampled the output of each layer in the decoder to the same size of the original image, and then used the multi-scale method for auxiliary training. In this paper, we adopt a multi-scale loss for the training as well. We resize the labels to the same size of each layer of the decoder to formulate the objective function. The ground-truth is then resized to the same size of the corresponding layer. Let $\mathcal{W}$ be the parameters of the convolutional layers,   $N$ as the number of decoder layers and    $w=(w_1, w_2,\dots, w_N)$ as the corresponding weights. The multi-scale loss is computed as follows.
\begin{equation}
\mathcal{L}\left(\mathcal{W},w\right)=\sum_{n=1}^{N}{\lambda_{n}L^{(n)}\left(\mathcal{W},w_{n}\right)},
\end{equation} 
where $\lambda_{n}$ is the weight of $n^{th}$ layer in decoder, $L^{(n)}$ represents the loss of the $n^{th}$ layer. In our paper, we use equal weights as that in \cite{fu2018joint} and set $\lambda_{n}=1/N$. 

The use of the loss function is flexible and can be a  {mean square} error (MSE) loss, a cross entropy (CE) loss or a Dice loss. It is also possible to apply different loss functions to different layers. We observed that there is a data imbalance between the foreground and the background in the vessel segmentation task. Recent work \cite{vnet} and \cite{sudre2017generalised} have shown that the Dice coefficient can effectively reduce the negative  {effect} of data imbalance.  Thus, we choose the Dice coefficient for the final loss computation.  In our paper, the multi-scale Dice coefficient loss $L_{dice}^{(n)}$ is computed as:
\begin{equation}
L_{dice}^{(n)}=1-\frac{2\sum_{i=1}^{K}p_i g_i+\epsilon}{\sum_{i=1}^{K}p_i^2+\sum_{i=1}^{K}g_i^2+\epsilon},
\label{mseloss}
\end{equation}
where $K$ denotes the number of pixels, $p_i \in [0,1]$ and $g_i \in \{0,1\}$ denote the predicted probability and ground truth value of $i^{th}$ pixel, respectively. The parameter $\epsilon$ ($\epsilon=1.0$ in this paper) is a Laplace smoothing factor used to avoid numerical problems and accelerate the convergence of the training process. We set $N=4$ in  {this work}, which means that the encoder and decoder have four down-sampling layers and up-sampling layers, respectively. The gradient of $L_{dice}^{(n)}$ with respect to $p_i$ is expressed as:
\begin{equation}
\begin{aligned}
\frac{\partial L_{dice}^{(n)}}{\partial p_i}=&-\frac{2 g_i}{\left(\sum_{i=1}^{K}p_i^2 + \sum_{i=1}^{K}g_i^2 +\epsilon\right)}\\
&+\frac{2 p_i \left(2 \sum_{i=1}^{K}p_i g_i + \epsilon \right)}{\left( \sum_{i=1}^{K}p_i^2 + \sum_{i=1}^{K}g_i^2 +\epsilon \right)^2}.
\end{aligned}
\end{equation}
Then we use the standard stochastic gradient descent for back-propagation and   the optimization.
% 
%%%%%%%%%%%%%%%%%%%%%%%%%%%%%%%%%%%%%%%%%%%
%\begin{figure}[htb]
%\centering
%\includegraphics[width=0.35\textwidth]{heatmap.png}
%\caption{Diagram of the probability map $P_{nn}$. The probability map $P_{nn}$ is a heatmap that reflects the probability of every pixel.}
%\label{heatmap}
%\end{figure}
%%%%%%%%%%%%%%%%%%%%%%%%%%%%%%%%%%%%%%%%%%%
%%%%%%%%%%%%%%%%%%%%%%%%%%%%%%%%%%%%%%

\section{Probability Regularized Walk}
Although many deep learning based methods have been proposed for retinal vessel segmentation, most of the algorithms did not consider the  {structural} connectivity. Therefore, fractured vessel segments often appear in the results \cite{zhang2018reconnection}. The broken vessel segmentation often post challenge to the subsequent analysis of the vessels such as the topology, the tortuosity, vessel lengths, etc. To obtain a better geometric description of the vessels, we propose a PRW algorithm to connect the broken vessel segments. 

Random walk is one of the most widely known and used method in graph theory \cite{lovasz1993random}. Let $G = (V, E)$ denote an undirected graph with a set of vertices $V$ and a set of edges $E$. A random walk  {process} in such graph can be characterized by  {a series of} transition probabilities  {among} its vertices. Many early work applied random walk to semantic segmentation \cite{grady2006random} \cite{li2015automated} \cite{dong2016sub}. These methods are often based on the similarity of  {pixels in the neighbour}. In deep learning, we often compute the probabilities of pixels belonging to foreground or background, which are  {binarized} with threshold to get the vessel map. Clearly, the binarization may discard some information such as the difference between the probabilities and the threshold values. Inspired from that, {PRW} integrates the probability outputs of deep learning into the regularized walk. In the vessel reconnection tasks, we treat each pixel in the image as a separate node in the graph, where the similarity between the two nodes can be evaluated based on the DDNet prediction probability. Since PRW aims to connect broken vessels, we expect the broken vessels  {to} lie along the direction of the remaining vessels. Therefore, we integrate the local vessel directions as well. It should be pointed out that we only borrowed and simulated the walking process of the walker in the PRW instead of the random walk algorithm in the strict sense.

The PRW takes the  {probability} map of DDNet and the binary vessel map as two inputs. Firstly, we label all the connected components in the binary map using morphological operations (bwlabel function in Matlab). For simplicity, we  {denote} the largest connected component as $L_1$ and the rest of components as $L_i\;(i=2,\dots,N_c)$, where $N_c$ is the number of components. Next, we connect each $L_i$ to $L_1$ iteratively to obtain the final vessel map. 

\begin{figure} 
	\centering
	\includegraphics[width=0.48\textwidth]{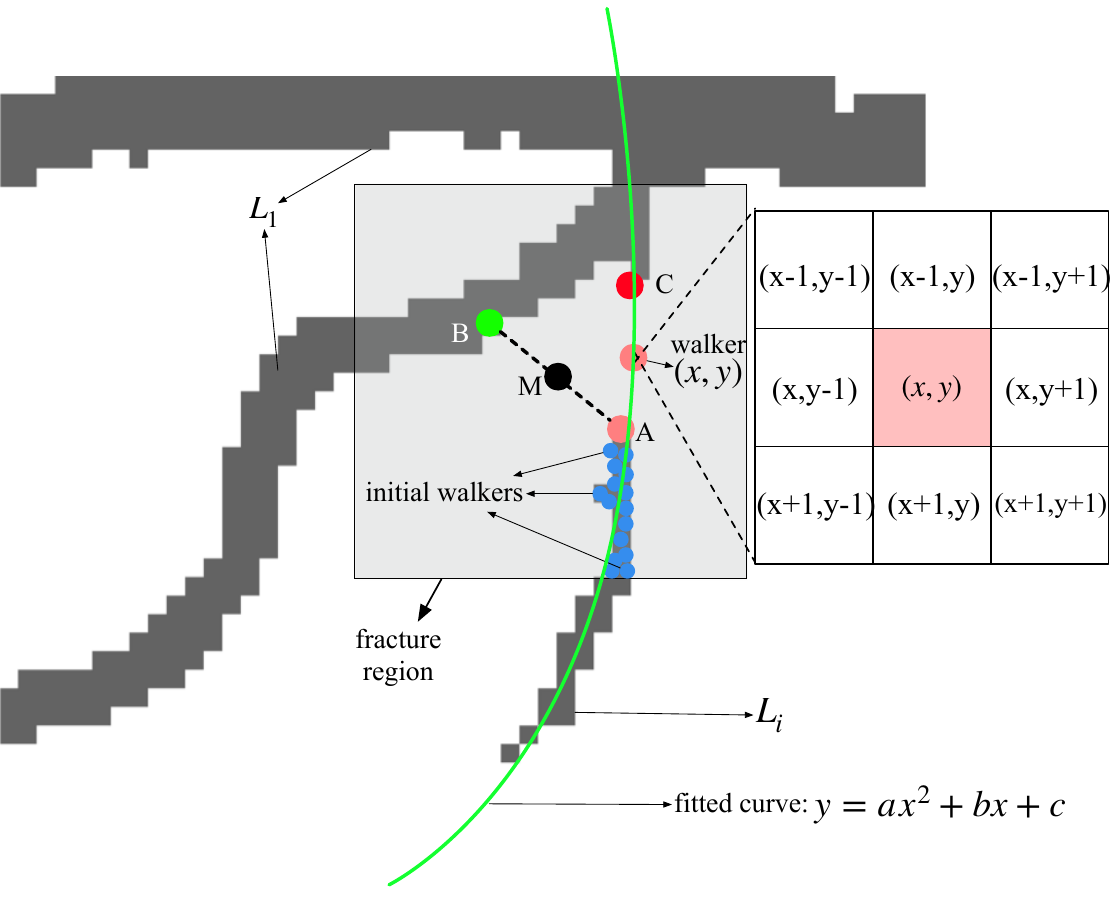}
	\caption{Illustration of regularized walk. Grey area represents fracture region. In the fracture region, the pink point represents the walker with its position $(x,y)$ during the migration. $A$ is the ending point of $L_{i}$, and $B$ is the closest point to $A$ in $L_{1}$, and $M$ is the middle point of $A$ and $B$. The blue dots at the intersection of $L_{i}$ and the fracture region indicate the initial location of the walkers. The green line represents the curve that is fitted by points in $L_{i}$. The square on the right side of the figure represents the 8-connected neighbors of the walker (shown in pink), and the numerical values represent the coordinates.}
	\label{randomwalk}
\end{figure}
%%%%%%%%%%%%%%%%%%%%%%%%%%%%%%%%

To connect $L_i$ to $L_1$, we first detect the fractured vessels between $L_i$ and $L_1$. As illustrated in Fig. \ref{randomwalk}, two points $A$ and $B$ with minimum distance in between are found from $L_{i}$ and $L_{1}$.  {Mathematically}, 
\begin{equation}
d_{AB}=\mathop{\arg\min}_{\mbox{\tiny$\begin{array}{c}\left(x_{A},y_{A}\right)\in L_{i}\\
\left(x_{B},y_{B}\right) \in L_{1}\end{array}$}} \sqrt{\left(x_{A}-x_{B}\right)^{2}+\left(y_{A}-y_{B}\right)^{2}}.
\end{equation}
where $d_{AB}$ denotes the distance between $A$ and $B$. Note that we calculate the $d_{AB}$ only using the centerline pixels of $L_i$ and $L_1$. Let $L_i$ and $L_1$ contain $m$ and $n$ pixels ($m \ll n$), respectively. Thus, the time complexity of the operation is: $O(m\times n)$.

In the second step, we determine a region of interest (ROI) where possible missing vessels  {are detected}. Our initial method of detecting ROI is to find all fractured and isolated vessels by morphological operation. Denote the center point between $A$ and $B$ as $M$ and a $l\times l$ square centered at $M$ is determined as the ROI. In this paper, we set $l=100$ empirically. With this threshold, we do not connect $L_i$ too far away from $L_1$, i.e., $d_{AB}>l$. 

In the third step, we calculate the main walking direction of the walker. Based on the observation that the missing vessels are usually on the same line of the remaining vessels $L_{i}$, we fit the vessel using polynomial curves. In this paper, a second order polynomial $y=ax^{2}+bx+c$  {is adopted}. All  {the} points in $L_{i}$ are used to fit the curve and determine the polynomial function. Then the intersecting point between the polynomial curve and $L_{1}$ is determined as $C$ with coordinates of $(x_c, y_c)$. Fig. \ref{randomwalk} illustrates the process.

In the fourth step, we use all points that belong to both $L_i$ and ROI as the initial walkers (seed pixels) and let these points walk iteratively until they arrive $L_1$. We optimize  {this} walking process by the predicted probability of the neural network ($P_{nn}$) and the direction probability ($P_d$) of the current walker. The detailed description of $P_{nn}$ and $P_d$ is as follows.

An important issue in the fourth step is the walking direction in each movement. Intuitively, we hope the walker to  {move} toward point $C$. However, as discussed, the binarization discards some information and a straight line or a fitted curve may discard the local changes. Therefore, we propose to regularize the walking direction using the probability map from the DDNet. The probability of a walker at $(x,y)$ walking toward $(x', y')$ is computed as $ P(x',y')$:
\begin{equation}
P\left(x',y'\right)=\alpha P_d\left(x',y'\right)+\left(1-\alpha\right)P_{nn}\left(x',y'\right), \label{eq5}
\end{equation}
where the first item $ P_d(x',y')$ is a probability which denotes the association between the main walking direction and the direction from $(x,y)$ to $(x',y')$.  The second item $ P_{nn}(x',y')$ denotes the probability output of DDNet. It indicates the probability of the pixel at $(x',y')$ being a vessel pixel. $\alpha$ controls the trade-off between the two items. 

In order to let the walker walks toward $C$, we design $P_d$ such that a pixel with smaller distance to $C$ has larger probability. For a pixel $(x',y')$ where $x'\neq x_c$ or $y'\neq y_c$, we compute $P_d$ as
\begin{equation}
P_d(x',y')=\frac{1}{\sqrt{\left(x'-x_{c}\right)^{2}+\left(y'-y_{c}\right)^{2}}},
\end{equation}
We set $\alpha=0.2$ empirically in (\ref{eq5}) such that neither item will dominate the results. In each step, we let the walker walk toward its 8-connected neighbor with largest probability $P(x',y')$:
\begin{equation}
\label{max-loc}
\left(x',y'\right)=\mathop{\arg\max}_{(x',y')\in \Omega} \; \{P\left(x',y'\right)\},
\end{equation}
where $\Omega$ is consisted of the 8-connected neighbors of $(x,y)$, i.e., $\Omega =\{ (x-1,y-1),(x-1,y),(x-1,y+1),(x, y-1),(x, y+1),(x+1,y-1),(x+1,y),(x+1,y+1)\}$. Additionally, if there is always $P_{nn}<\epsilon_{nn}$ ($\epsilon_{nn}$ is set to be 0.1 in this paper) around the walker in the walking process, we stop the movement of the walker and determine that the current ROI is not a fractured vessel or the terminal point  $A$  we selected is  {in} the wrong direction. When point $A$ is not the right point for  {connecting}, the corresponding probability at the point will be low. In this case, the algorithm will  {terminate the} connect {operation} to $L_1$  {to} avoid mis-connection. Meanwhile, it may still have the chance
to be connected to $L_1$ as the walker can walk from point $C$ in
$L_1$. In order to obtain the newly created vessels without holes, we use a $3\times 3$ kernel to simulate walkers during walking process. When all the walker activities are completed, there are many paths established in the ROI, or there is no path (indicating that the two regions are not expected to be connected). For time complexity, assume that the DDNet prediction binary map contains $k$ fracture regions (ROI). Within each ROI, the PRW requires a maximum of $l\times l$ loops to traverse each pixel to reconstruct the blood vessel, where $l$ is the length of the ROI. Therefore, the time complexity of the PRW is $O(k\times l^2)$.

\section{Experiments}
\label{experiments}
\subsection{Dataset}
\label{dataset}
We use the following three public datasets in our experiments: {DRIVE} \cite{drive}, STARE  \cite{stare}, and CHASE\_DB1\cite{chasedb1}.

The {DRIVE\footnote{\url{http://www.isi.uu.nl/Research/Databases/DRIVE/}}} dataset contains 40 color fundus images, which were originally divided into 20 images for training and 20 images for testing. The images were acquired using a Canon CR5 non-mydriatic 3-CCD camera with a $45^{\circ}$ field of view (FOV). Each image in this dataset has a dimension of $565 \times 584$. We follow the same partition of the images in our training and testing. 

The {STARE\footnote{\url{http://www.ces.clemson.edu/$\sim$ahoover/stare/probing/index.html}}} dataset comprises of 20 color fundus images. The images were captured by a Topcon TRV-50 fundus camera at $35^{\circ}$ FOV. Half of images contain pathological indications and the other half is from healthy subjects. Each image has a dimension of $700 \times 605$. However, unlike the {DRIVE} dataset, there is no fixed partition of training and testing sets for STARE. In this paper, we adopt the k-fold ($k=4$) cross-validation \cite{Fushiki2011} for the training and testing phases, similar to that in \cite{Zhang2018}. Therefore, we have 15 images for training and the remaining 5 images for testing in each fold.

The {CHASE\_DB1\footnote{\url{http://blogs.kingston.ac.uk/retinal/chasedb1/}}} dataset contains 28 images from both the left and right eyes, and the images were captured with a $30^{\circ}$ FOV. Each image has a dimension of $999 \times 960$. For convenience, we   {still} use the cross-validation method as STARE. In each fold, 21 images are used for training and the rest of 7 images are used for testing.

 {Considering that } there are two sets of manual ground truth provided for the datasets, we follow \cite{fu2016deepvessel} and \cite{lin2018automatic} to use the manual annotation from the first observer as the ground truth for all the images.

\subsection{Evaluation Metrics}
In order to quantitatively evaluate our method, we compute Specificity (Spe), Sensitivity (Sen) and Accuracy (Acc) as our evaluation metrics:
\begin{equation}
Spe=\frac{TN}{TN+FP}, 
\end{equation}
\begin{equation}
Sen=\frac{TP}{TP+FN},
\end{equation}
\begin{equation}
Acc=\frac{TP+TN}{TP+FP+TN+FN},
\end{equation}
where TP, FN, TN, and FP represent true positive, false negative, true negative and false positive, respectively. In addition, we also introduce the area under receiver operating characteristic (ROC) curve (AUC) metric.

To evaluate the performance of the PRW algorithm in the fracture region, we define a new metric $Err$ to explain how the PRW reconnects the vessel successfully:
\begin{equation}
Err=\frac{1}{N_r}\sum_{i=1}^{N_r}\left(\frac{FP}{TP+FP}\right)^{(i)},
\label{err}
\end{equation}
where  $\left(\frac{FP}{TP+FP}\right)^{(i)}$ denotes the pixel error of the $i^{th}$ ROI detected by PRW, and $N_r$ is the number of ROIs. The metric $Err$ explains the error rate when the regularized walk algorithms are used to recover broken vessels. A smaller $Err$ indicates a better performance of the regularized walk algorithm.

%%%%%%%%%%%%%%%%%%%%%%%%%%%%%%%%%%%%%%%%
\begin{table*}[t]
\centering
\caption{$Acc$, $Sen$ and time consumption of PRW at different ROI thresholds.}
\begin{tabular}{c|c|c|c|c|c|c|c|c|c}\hline
Threshold & 0 & 20 & 40 & 60 & 80 & \textbf{100} & 120 &140 & 160 \\\hline
Acc & 0.9594 & 0.9594 & 0.9595 & 0.9599 & 0.9603 & \textbf{0.9605} & 0.9605 & 0.9605 & 0.9605 \\\hline
Sen & 0.8126 & 0.8126 & 0.8127 & 0.8128 & 0.8130 & \textbf{0.8131} & 0.8131 & 0.8131 & 0.8131 \\\hline
Time(s) & - & - & 3.786 & 20.695 & 54.425 & \textbf{81.308} & 117.128 & 164.363 & 252.148 \\\hline 
\end{tabular}
\label{tab:thresh-time}
\end{table*}

\begin{table*}[h]
\centering
\caption{Error rates ($Err$) of PRW at different $\alpha$ values.}
\begin{tabular}{c|c|c|c|c|c|c|c|c|c|c|c}\hline
$\alpha$ & 0.0 & 0.05 & 0.10 & 0.15 & 0.20 & 0.25 & 0.30 & 0.35 & 0.40 & 0.45 & 0.50 \\\hline
$Err$ & 0.0 & 0.1145 & 0.1023 & 0.0197 & \textbf{0.0173} & 0.0188 & 0.1043 & 0.1169 & 0.2031 & 0.2165 & 0.2327 \\\hline
\end{tabular}
\label{tab:alpha}
\end{table*}
%%%%%%%%%%%%%%%%%%%%%%%%%%%%%%%%%%%%%%%%

\subsection{Training and Testing Phases} 
We implemented DDNet on PyTorch framework with a single NVIDIA GPU (GeForce GTX 1080). During the training, we employ adaptive moment estimation (Adam) for optimizing the deep model. The initial learning rate is set to $0.001$ and a weight decay of $0.0005$. We follow the same poly learning rate policy in \cite{zhao2017pyramid} and the maximum epoch is 1000. Due to GPU memory limitations, we set the batch size to 6 throughout the training. All training images are resized to $512\times 512$. The output of the DDNet is a 1-channel map. We use the OTSU thresholding method \cite{otsu} to binarize the output after the testing. PRW algorithm is implemented on the Matlab platform. The PRW algorithm takes the output of the DDNet as input and  {delivers} the binary map of reconnected vessels.

We perform data  {augmentation} on the training and validation images with a random rotation of an angle from $-45^{\circ}$ to $45^{\circ}$ and a random contrast enhancement in the range $[-2, 2]$. The labels of each dataset are resized to $256\times256$, $128\times128$, $64\times64$ during the training. In the testing phase, the input image is resized to $512\times512$, without any enhancements. It is worth pointing out that the proposed DDNet reached an average inference time of 145.61ms with a total of 56.03M parameters and a total of 476.08G FLOPs in the testing phase.

\subsection{Hyperparameters Setting}
\label{param}
We introduce two hyperparameters $l$ and $\alpha$ in the PRW algorithm, which control the detection range of the ROI and the movement behavior of the walkers. For ROI detection, a small threshold causes the ROI to fail to cover the fracture region, and an excessive threshold increases the time complexity. During the movement of walkers, an excessive $\alpha$ will cause the force of $P_d$ to be much larger than $P_{nn}$, thus causing the walker to break away from the real trajectory. Therefore, the selection of appropriate hyperparameters will promote the performance and efficiency of vascular reconnection.

%%%%%%%%%%%%%%%%%%%%%%%%%%%%%%%%%%%%%%%%%%%%%
\begin{figure} 
	\centering
	\includegraphics[width=0.48\textwidth]{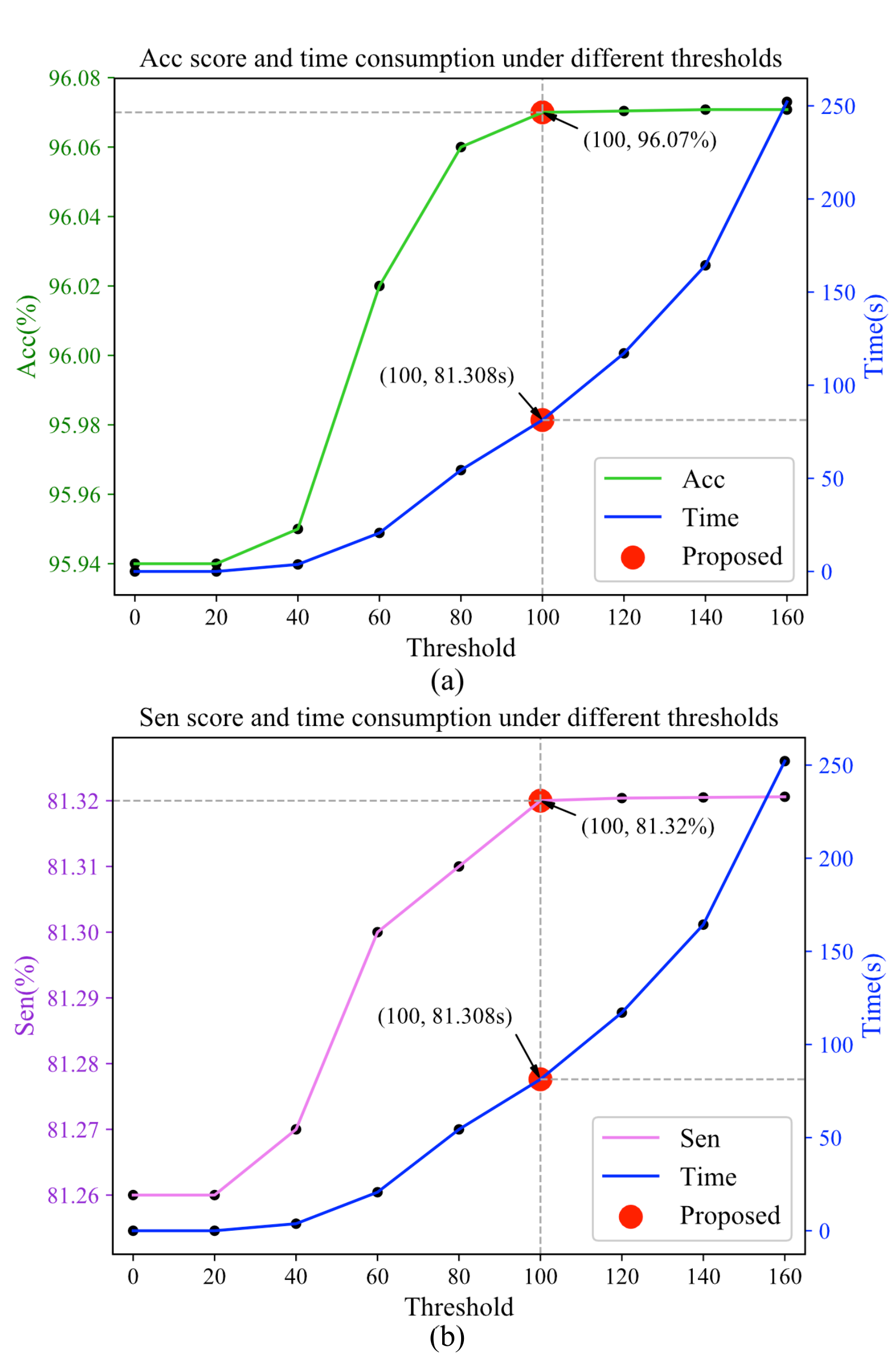}
	\caption{$Acc$ and $Sen$ scores and time consumption results at different thresholds. (a) The green line with a left vertical axis indicates the $Acc$ scores at different thresholds, and the blue line with a right vertical axis shows the time consumption of PRW at different thresholds. (b) The pink line with a left vertical axis indicates the $Sen$ scores at different thresholds and the blue line with a right vertical axis shows the time consumption of PRW at different  {thresholds. The} red point indicates the $Acc$ and $Sen$ score and time consumption under the threshold in this paper.}
	\label{thresh-time}
\end{figure}
%%%%%%%%%%%%%%%%%%%%%%%%%%%%%%%%%%%%%%%%%%%%%%

Firstly, we set a series of thresholds $l$ from 0 to 160 with a step of 20 and implement the PRW algorithm at different values. Then, we compute the $Acc$ and $Sen$ scores and time consumption (Time) for each set of experiments. The comparison results are shown in Fig. \ref{thresh-time} and Table \ref{tab:thresh-time}. As shown in the figure, further increasing the  {threshold} $l$ does lead to higher $Acc$ and $Sen$, though the time consumption is increased. Therefore, we set $l=100$. Secondly, we set a series of $\alpha$ from 0.05 to 0.5 with a step of 0.05 and implement PRW in each setting. We compute the reconnection error rate ($Err$) for different $\alpha$ values. The lower the $Err$, the better the performance of the algorithm. Table \ref{tab:alpha} shows the $Err$ of the PRW at different $\alpha$ values. From Table \ref{tab:alpha}, we can conclude that when $\alpha=0.2$, the error rate is the lowest. It should be explained that when $\alpha =0$, $Err=0$. The reason for this is that when $\alpha=0$, $P_d=0$, so that $P(x',y')=P_{nn}$. That is to say, the probability of the walker around the initial position is the largest, so that the walker will not be driven to escape from the initial position to reach $L_1$. Similarly, we reached the same conclusion on STARE and CHASE\_DB1.

It can be  {observed} from the first experimental result that a larger threshold $l$ enables the ROI to cover  {fracture region}. However, the number of fracture region is limited in the prediction map, and when $l>100$, the AUC score hardly increases. From the analysis of the second experimental result, we conclude that the vessel reconnection error rate is the lowest when $\alpha = 0.2$, and the inference time required for PRW increases exponentially when $\alpha > 0.2$. Based on the above two experimental results, we can conclude that the PRW algorithm performs best in the vessel reconnection task when $l=100$ and $\alpha=0.2$.

\subsection{Ablation Studies}
In the proposed method, we have introduced the ResNet blocks, the DDB, the multi-scale dice loss, and the PRW algorithm to improve the baseline encoder-decoder method. To justify the effectiveness of these modules, we conduct   the following ablation studies. For simplicity, the DRIVE dataset is used. We start from the baseline U-Net \cite{unet} approach and evaluate how these modules affect the results.

%%%%%%%%%%%%%%%%%%%%%%%%%%%%%%%%%%%%%%%%%%%%%%%
\begin{table}[t]
\centering
\caption{Ablation study for replacing original block with ResNet block}
\begin{tabular}{ccccc} \hline
%\toprule
Method & Acc & Sen & Spe & AUC  \\\hline
%\midrule
U-Net & 0.9485 & 0.7356 & 0.9602 & 0.9641 \\ \hline
ResUNet & \textbf{0.9501} &  \textbf{0.7887} &  \textbf{0.9632} &  \textbf{0.9665}  \\\hline
%\bottomrule
\end{tabular}
\label{ab:res}
\end{table}
%%%%%%%%%%%%%%%%%%%%%%%%%%%%%%%%%%%%%%%%%%%%%%%

\subsubsection{Ablation study for replacing original blocks with ResNet blocks}
To justify the use of the ResNet blocks, we replace the original blocks in the U-Net with ResNet blocks. We denote the modiﬁed U-Net as ResUNet. Table \ref{ab:res} shows the segmentation results. Our results show that we achieve 0.9501, 0.7887, 0.9632, and 0.9665 on Acc, Sen, Spe and AUC in comparison to 0.9485, 0.7356, 0.9602,  {and} 0.9641 in the original U-Net.

%%%%%%%%%%%%%%%%%%%%%%%%%%%%%%%%%%%%%%%%%%%%%%%%
\begin{table}[t]
\centering
\caption{Ablation study for the dense dilated block.}
\begin{tabular}{ccccc}\hline
%\toprule
Method & Acc & Sen & Spe & AUC \\\hline
%\midrule
ResUNet & {0.9501} &  {0.7887} &  {0.9632} &  {0.9665} \\ \hline
ResUNet + Cascade Mode & 0.9578 &  0.8109 & 0.9780 &  0.9781 \\\hline
ResUNet + Parallel Mode & 0.9564 &  0.8064 &  0.9783 &  0.9776 \\\hline
ResUNet + Dense Mode &  \textbf{0.9594} & \textbf{0.8126} & \textbf{0.9788} & \textbf{0.9796}\\\hline
%\bottomrule
\end{tabular}
\label{ab:ddb}
\end{table}
%%%%%%%%%%%%%%%%%%%%%%%%%%%%%%%%%%%%%%%%%%%%%%%%%%

%%%%%%%%%%%%%%%%%%%%%%%%%%%%%%%%%%%%%%%%%%%%%%%
\begin{table}[t]
\centering
\caption{Ablation study for multi-scale dice loss.}
\begin{tabular}{ccccc}\hline
%\toprule
Method & Acc & Sen & Spe & AUC \\\hline
%\midrule
DDNet + CE &  0.9561 & 0.8089 & 0.9781 & 0.9780 \\\hline
DDNet + MSE & 0.9573 & 0.8100 & 0.9775 & 0.9784 \\\hline
\textbf{DDNet + M-net Loss} &0.9590 & 0.8121 & 0.9779 & 0.9789 \\\hline
DDNet + MSD & \textbf{0.9594} & \textbf{0.8126} & \textbf{0.9788} & \textbf{0.9796}\\\hline
%\bottomrule
\end{tabular}
\label{ab:msll}
\end{table}
%%%%%%%%%%%%%%%%%%%%%%%%%%%%%%%%%%%%%%%%%%%%%%%%%%
%%%%%%%%%%%%%%%%%%%%%%%%%%%%%%%%%%%%%%%%%%%%%%%%%%%
\begin{table}[t]
\centering
\caption{Ablation study for probability regularized walk.}
\begin{tabular}{cccc}\hline
%\toprule
Method & Acc & Sen & Spe \\\hline
%\midrule
DDNet & 0.9594 & 0.8126 & \textbf{0.9788}\\\hline
DDNet + CRW & 0.9601 & 0.8128 &  0.9782\\\hline
DDNet + PRW & \textbf{0.9607} & \textbf{0.8132} & 0.9783 \\\hline
%\bottomrule
\end{tabular}
\label{ab:rw}
\end{table}

\begin{table}[h] 
	\caption{The  pixel errors ($Err$) by CRW and PRW on DRIVE, STARE and CHASE\_DB1 datasets.}
	\centering
	\begin{tabular}{c|c|c|c}\hline
		%\toprule
		\multirow{2}*{Methods} &\multicolumn{1}{c}{DRIVE} & \multicolumn{1}{c}{STARE} & \multicolumn{1}{c}{CHASE\_DB1} \\
		\cline{2-4}
		~ &$Err$ &$Err$ &$Err$\\
		\hline
		DDNet+CRW & 0.1390 & 0.1286 & 0.1054 \\
		\hline
		DDNet+PRW & \textbf{0.0173} & \textbf{0.0195} & \textbf{0.0160}\\\hline
		%\bottomrule
	\end{tabular}
	\label{tab:conn}
\end{table}

%%%%%%%%%%%%%%%%%%%%%%%%%%%%%%%%%%%%%%%%%%%%%%%%%%%
\begin{table*}[t] 
\centering
\caption{Performance comparison on the DRIVE, STARE and CHASE\_DB1 datasets. }
\begin{tabular}{c|cccc|cccc|ccccccc}\hline
%\toprule
\multirow{2}*{Methods} & \multicolumn{4}{c}{DRIVE}&\multicolumn{4}{c}{STARE}&\multicolumn{4}{c}{CHASE\_DB1}  \\
\cline{2-13}
 ~& Acc&Sen&Spe&AUC&Acc&Sen&Spe&AUC&Acc&Sen&Spe&AUC \\ 
\hline
2nd Human Observer & 0.9472&0.7760&0.9724&-&0.9349&0.8952&0.9384&-&0.9545&0.8105&0.9711&- \\
\hline
Azzopardi \cite{azzopardi2015trainable} & 0.9442 & 0.7655 & 0.9704 & 0.9614 & 0.9497 & 0.7716 & 0.9701 &0.9563& 0.9387 & 0.7585 & 0.9587 &0.9487 \\\hline
Ronneberger \cite{unet} &  0.9501 & 0.7356 & 0.9602 & 0.9641 & 0.9517 & 0.7101 & 0.9682 & 0.9615 & 0.9499 & 0.7094 & 0.9767 & 0.9613 \\\hline
Zhao \cite{zhao2015automated} & 0.9540 & 0.7420 & \textbf{0.9820} & 0.8620 & 0.9560 & 0.7800 & 0.9780 & 0.8740 & - & - & - & - \\\hline
Roychowdhury \cite{roychowdhury2015iterative} & 0.9494 & 0.7395 & 0.9782 &0.9672& 0.9560 & 0.7317 & \textbf{0.9842} &0.9673& 0.9467 & 0.7615&0.9575 &0.9623 \\\hline
Zhao \cite{zhao2018automatic}& 0.9580 & 0.7740 & 0.9790 & 0.9750 & 0.9570 & 0.7880 & 0.9760 & 0.9590 &-&-&-&-\\ 
\hline
Xie \cite{xie2015holistically} & 0.9435 & 0.7364 &0.9730 &0.9774& 0.9402 & 0.7116 &0.9724 &0.9801& 0.9380 & 0.7151 & 0.9679 &0.9798 \\\hline
Fu \cite{fu2016deepvessel} & 0.9533 & 0.7603 & 0.9776 &0.9789& 0.9609 & 0.7412 & 0.9701 &0.9790& 0.9581 & 0.7130 & \textbf{0.9812} &0.9806 \\\hline
Zhou \cite{zhou2017improving} & 0.9469 &  0.8078 &  0.9674 & - & 0.9585 &  0.8065 & 0.9761 & - &  0.9520 &  0.7553 & 0.9751 & - \\\hline
\textbf{Our DDNet} & \textbf{0.9594} & 0.8126 & 0.9788 & \textbf{0.9796} & \textbf{0.9685} & \textbf{0.8391} & 0.9769 & \textbf{0.9858} & \textbf{0.9637} & \textbf{0.8268} & 0.9773 & \textbf{0.9812} \\\hline\hline
\textbf{Our DDNet+PRW}& 0.9607 & 0.8132 & 0.9783 & - & 0.9698 & 0.8398 & 0.9761 &-& 0.9648 & 0.8275 & 0.9768 &- \\\hline
%\bottomrule
\end{tabular}
\label{tab:result}
\end{table*}

%%%%%%%%%%%%%%%%%%%%%%%%%%%%%%%%%%%%%%%%%%%%%%%%%%%
\begin{figure}[t]
	\centering
	\includegraphics[width=0.45\textwidth]{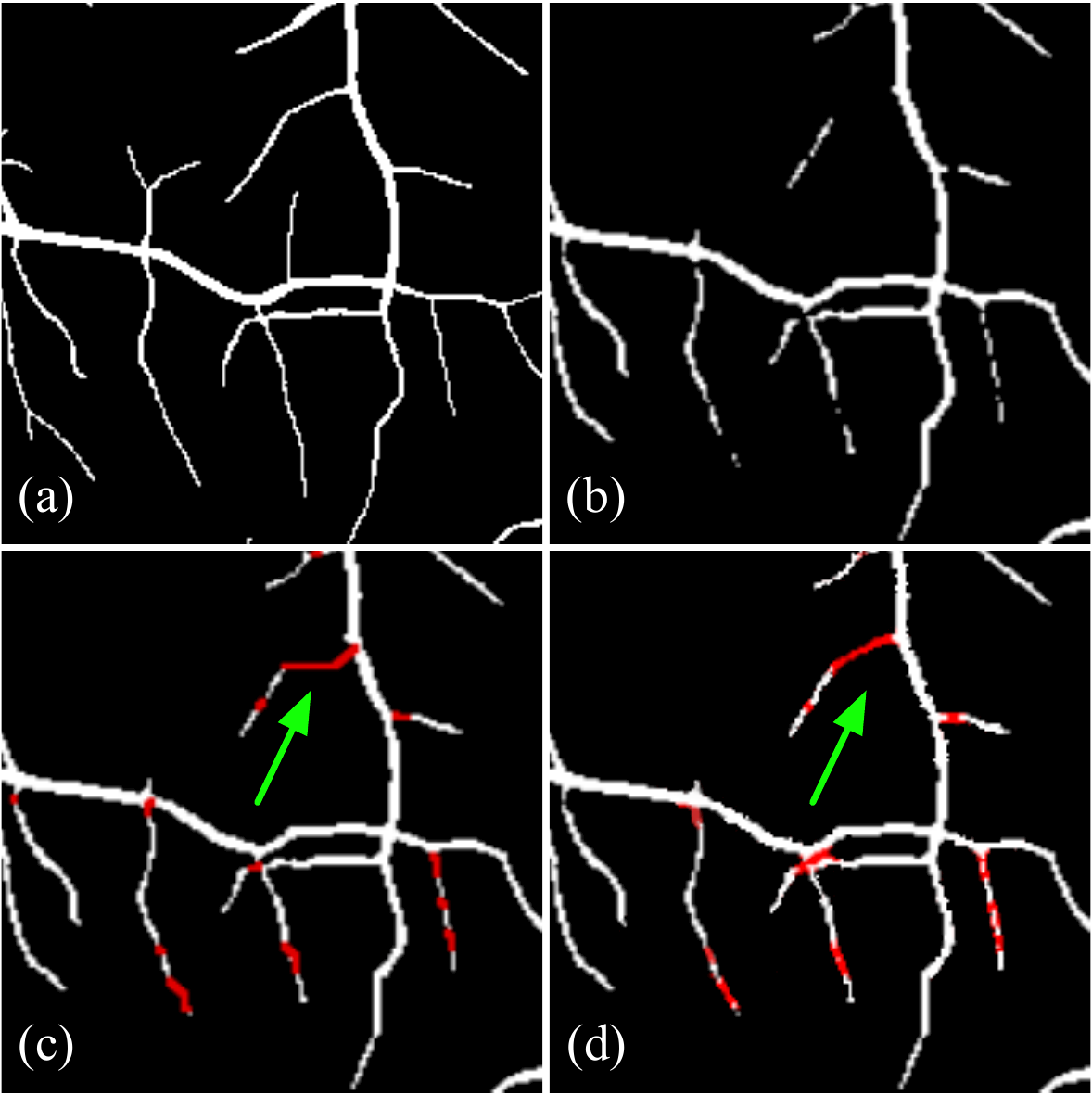}
	\caption{Illustration of vessel reconnection. (a) manual ground truth. (b) result of DDNet. (c) DDNet+CRW. (d) DDNet+PRW.}
	\label{diffrw}
\end{figure}
%%%%%%%%%%%%%%%%%%%%%%%%%%%%%%%%%%%%%%%%%%%%%%%%%%%

\subsubsection{Ablation study for dense dilated block}
To justify the use of the proposed DDB module, we compare the dense mode with cascade mode, parallel mode as well as the ResUNet without additional blocks. The ResUNet is  {taken} as baseline and the multi-scale dice loss is used in all experiments. For fair comparison, we set up 12 dilated convolutional layers in cascade mode and the parallel mode. The dilation rates are set to 1, 2 and 5, as shown in Fig. \ref{dilatedconv}. The comparison results are shown in Table \ref{ab:ddb}. The results show that our DDB module is effective to improve the results and it performs better than the other two modes as well.

\subsubsection{Ablation study for multi-scale dice loss}
To justify the use of multi-scale dice loss (MSD), we  {compare} it with cross entropy loss (CE) and mean square error loss (MSE). Additionally, we set up another comparative experiment comparing the multi-scale loss in M-Net \cite{fu2018joint} (named as M-net Loss) with the proposed multi-scale label loss. In this ablation setting, the backbone is our proposed DDNet. We show the comparison results in Table \ref{ab:msll}. The results show that the multi-scale dice loss achieves highest AUC score of 0.9891. Therefore, the use of dice loss in the vessel segmentation task contributes to the performance improvement.
%%%%%%%%%%%%%%%%%%%%%%%%%%%%%%%%%%%%%%%%%%%%%%%%%%%

%%%%%%%%%%%%%%%

\subsubsection{Ablation study for probability regularized walk}
To justify the benefits of the proposed PRW, we compare the PRW with conventional walk (CRW). In addition, the results without any random walk are also given. All results are shown in Table \ref{ab:rw}. The results show that PRW leads to highest Acc and Sen  at 0.9607 and 0.8132, respectively. Since the PRW algorithm may misclassify some true negative (TN) pixels, $Spe$ has a 0.08\% decrease. Table \ref{tab:conn} compares the $Err$ by CRW and PRW in the DRIVE, STARE, and CHASE\_DB1 datasets. We can observe a large drop of the error $Err$ from CRW to PRW. We can conclude that the PRW effectively reconnects the interrupted region.
%%%%%%%%%%%%%%%%%%%%%%%%%%%%%%%%%%

Fig. \ref{diffrw} shows an example of detected vessels before and after the PRW as well as that after the CRW. In the figure, the red indicates the broken vessels recovered by  {both} algorithms. The green arrow indicates a scenario where the proposed PRW method works well while the CRW recovers the vessels incorrectly.

\subsection{Comparison With State-of-the-art Methods}
To justify the performance improvement compared with the state-of-the-art algorithms, we finally compare our method with other methods proposed by Azzopardi \cite{azzopardi2015trainable}, Zhao \cite{zhao2015automated}, Roychowdhury \cite{roychowdhury2015iterative}, Zhao \cite{zhao2018automatic}, Xie \cite{xie2015holistically}, Fu \cite{fu2016deepvessel}, Zhou \cite{zhou2017improving} and Ronneberger \cite{unet}. For the DRIVE dataset, the results are from the original papers. For STARE and CHASE\_DB1, the results are obtained by ourselves, where the original implementations of \cite{xie2015holistically} and \cite{fu2016deepvessel} are used to obtain the  {results} for Xie's method and Fu's method. For \cite{unet}, we replace the CE loss with the Dice loss from the original U-Net code as Dice loss works better. For \cite{zhou2017improving}, we implement the code and carefully fine-tune the parameters to get the results. As we mentioned in Section \ref{dataset}, we use four-fold cross-validation for the STARE and CHASE\_DB1 datasets. 
\begin{figure*}  
	\centering
	\includegraphics[width=0.95\textwidth]{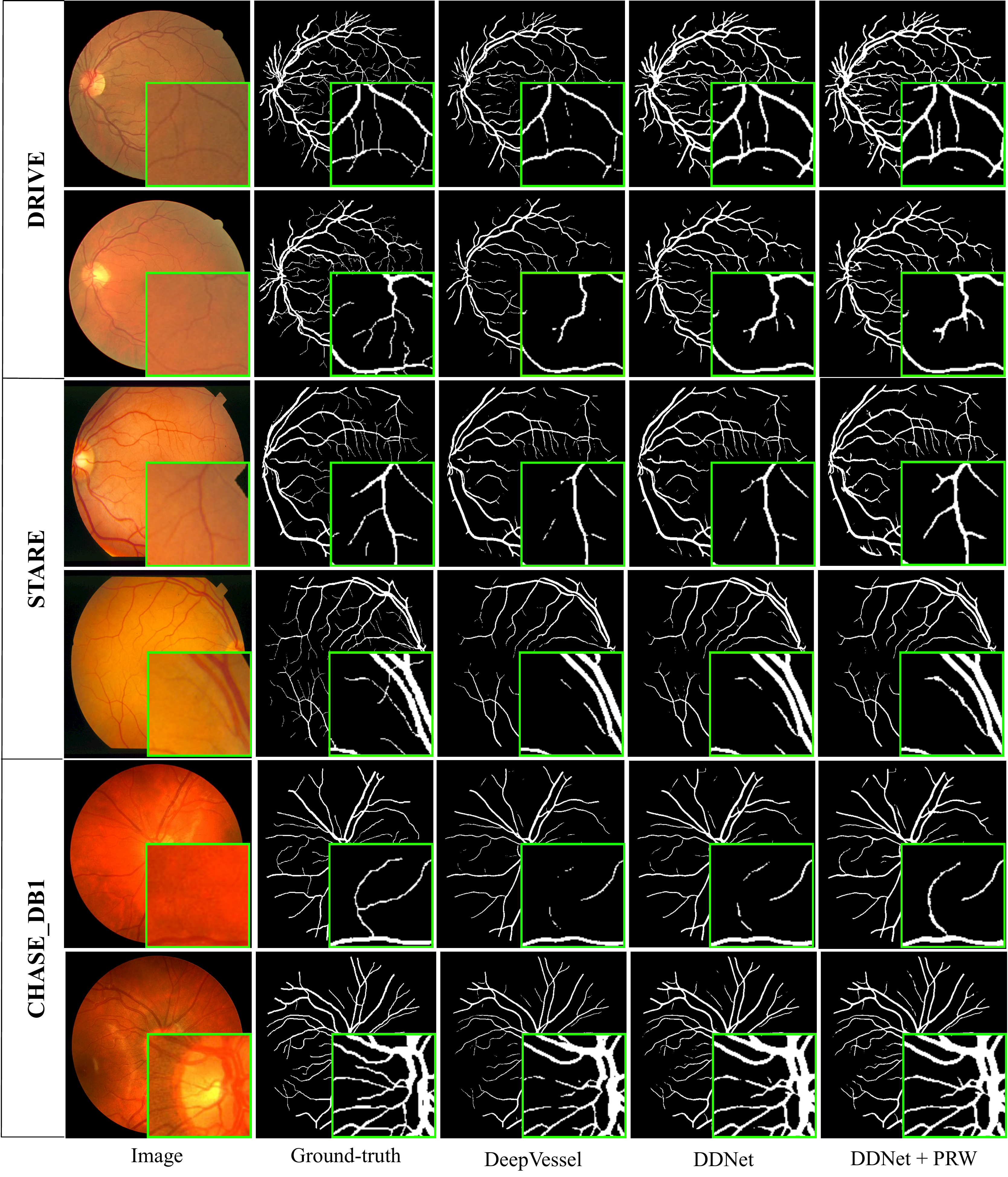}
	\caption{Visual illustration of segmentation results on DRIVE, STARE, CHASE\_DB1. From left to right: original image, ground-truth, results by DeepVessel \cite{fu2016deepvessel}, DDNet and DDNet + PRW. We show a local enlargement with a green box.}
	\label{visual_example}
\end{figure*}

Table \ref{tab:result} shows the $Acc$, $Sen$, $Spe$ and $AUC$ of these methods, the DDNet and the DDNet with PRW on the three datasets. From Table \ref{tab:result}, the proposed DDNet and PRW improve the accuracy of vessel segmentation consistently in all three datasets. 
 Fig. \ref{visual_example} shows results from three examples.  {From the figure}, the proposed method using DDNet and PRW segments vessels more accurately.

\section{Conclusions}
In this paper, we propose a dense dilated convolution network and a probability regularized walk algorithm to segment vessels and reconnect fractured vessels, respectively. The proposed DDNet integrates the context information of different layers and utilizes multi-scale dice loss to improve the segmentation of vessels. Dense Dilated Block (DDB) effectively improves the feature representation of deeper convolutional layers. And the efficient reuse of deep features  {facilitates efficiency of parameter usage}. The probability regularized walk algorithm effectively reconnects the fractured blood vessels, which makes the segmentation results better. The experiments show that the proposed method is able to achieve better vessel detection results than other methods. A limitation of our approach is that the PRW is proposed as a post-processing of DDNet. In future, we will explore to integrate the  {random} walk  {within} the deep learning.
\label{conclusions}
\bibliographystyle{IEEEbib}
\bibliography{references}
\end{document}